\documentstyle[12pt]{article}

\catcode`\@=11 \@addtoreset{equation}{section}\catcode`\@=12
\newcommand{\be}{\begin{equation}}
\newcommand{\ee}{\end{equation}}
\newcommand{\ba}{\begin{eqnarray}}
\newcommand{\ea}{\end{eqnarray}}

\begin{document}
\thispagestyle{empty}

\begin{center}
               RUSSIAN GRAVITATIONAL ASSOCIATION\\
               CENTER FOR SURFACE AND VACUUM RESEARCH\\
               DEPARTMENT OF FUNDAMENTAL INTERACTIONS AND METROLOGY\\
\end{center}
\vskip 4ex
\begin{flushright}                              RGA-CSVR-011/94\\
                                                gr-qc/9407028
\end{flushright}
\vskip 15mm

\begin{center}
{\large\bf  Billiard Representation for Multidimensional Cosmology
with Multicomponent Perfect Fluid near the Singularity}

\vskip 5mm
{\bf
V. D. Ivashchuk and V. N. Melnikov }\\
\vskip 5mm
     {\em Center for Surface and Vacuum Research,\\
     8 Kravchenko str., Moscow, 117331, Russia}\\
     e-mail: mel@cvsi.uucp.free.net \\
\end{center}
\vskip 10mm
ABSTRACT

The multidimensional cosmological model describing the evolution
of $n$ Einstein spaces in the presence of multicomponent perfect fluid is
considered. When certain restrictions on the parameters of the model are
imposed, the dynamics of the model near the singularity is reduced
to a billiard on the $(n-1)$-dimensional Lobachevsky space $H^{n-1}$.
The geometrical criterion for the finiteness of the billiard volume
and its compactness is suggested. This criterion reduces the problem
to the problem of illumination of $(n-2)$-dimensional sphere $S^{n-2}$
by point-like sources. Some generalization of the considered scheme
(including scalar field and quantum generalizations) are considered. \\

\vskip 10mm


\vskip 30mm

\centerline{Moscow 1994}
\pagebreak

\setcounter{page}{1}

\pagebreak

\section{Introduction}
\setcounter{equation}{0}

Last years multidimensional classical and quantum cosmology (see, for
example, [1-33] and references therein) became a rather popular object
of investigations both from physical and mathematical points of view.
A lot of interesting topics in multidimensional cosmology were considered:
exact solutions and problem of integrability, superstring cosmology
and problem of compactification, variation of constants,
classical and quantum wormholes, chaotic behavior near the singularity
etc.

In the present paper we deal with a stochastic behavior in multidimensional
cosmological models [24-33]. This direction in
higher-dimensional gravity was stimulated by well-known results for
"mixmaster" model [34-37]. We note, that there is also an elegant explanation
for stochastic behavior of scale factors of  Bianchi-IX model suggested by
Chitre [36,37] and recently considered
 in [38,39,40]. (For "history" of the problem see also [41].)
In the Chitre's approach the Bianchi-IX cosmology near the singularity
is reduced to a billiard on the Lobachevsky
space $H^2$ (see Fig. 4 below). The volume of this billiard is finite. This
fact together with the well-known behavior (exponential divergences) of
geodesics on the spaces of negative curvature leads to a stochastic
behavior of the dynamical system in the considered regime [42, 43].

It is quite natural to generalize the approach [36] to the multidimensional
case. This program was started in [33]. The present paper is devoted to
a construction of "billiard representation" for
the multidimensional cosmological model describing the evolution
of $n$ Einstein spaces in the presence of $(m+1)$-component perfect fluid
[21] (see Sec. 2). One of these components correspond to the cosmological
constant term [20]. In some sense the model [21] may be considered as
"universal" cosmological model: a lot of cosmological models
(not obviously multidimensional) may be embedded into this model.
(We  note also, that in some special cases the model  [21] was
considered by many authors [4,7,8,10,12,14-17]).

We impose certain restrictions on the parameters of the model [21] and
reduce its dynamics  near the singularity  to a billiard on the
$(n-1)$-dimensional Lobachevsky
space $H^{n-1}$ (Sec. 3). The geometrical criterion for the
finiteness of the billiard volume and its compactness is suggested.
This criterion reduces the considered problem to the
geometrical (or topological) problem of
illumination of $(n-2)$-dimensional  unit sphere $S^{n-2}$
by $m_{+} \leq m$ point-like sources located outside the sphere [45,46].
These sources correspond to the components with $(u^{(\alpha)})^2 > 0$
(Sec. 3). When these sources illuminate the sphere then, and only
then, the billiard has a finite volume and the cosmological model
possesses a stochastic behavior near the singularity. (We note,
that,  for cosmological and curvature terms
$(u^{(\alpha)})^2 < 0$ and these terms may be neglected near
the singularity). For the case of an infinite billiard  volume the
cosmological model has a Kasner-like behavior near the singularity [18].
When the minimally coupled massless scalar field is added into consideration,
the evolution
in time is bounded: $t > t_0$ and the limit  $t \rightarrow t_0$ corresponds
to the approach to the singularity. In this case the stochastic behavior
near the singularity is absent.

In Sec. 4 we illustrate the suggested approach on an example
of the Bianchi-IX cosmology. In Sec. 5 the Wheeler-DeWitt equation
for the considered model in a special time gauge is considered. Near the
singularity this equation has an approximate solution generalizing
that for Bianchi-IX  case [40]. This solution may be considered as
a starting point for the construction of the third-quantized
cosmological models in the vicinity of the singularity (for the
"mixmaster" case one of such models was considered in [40]).

\section{ The model}
\setcounter{equation}{0}

In this paper we consider a cosmological model describing
the evolution of $n$ Einstein spaces in the presence of $(m+1)$-component
perfect-fluid matter [21]. The metric of the model
\begin{equation} 
g=- \exp[2{\gamma}(t)] dt \otimes dt +
\sum_{i=1}^{n} \exp[2{x^{i}}(t)]g^{(i)},
\end{equation}
is defined on the  manifold 
\begin{equation}
M = R \times M_{1} \times \ldots \times M_{n},
\end{equation}
where the manifold $M_{i}$ with the metric $g^{(i)}$ is an
Einstein space of dimension $N_{i}$, i.e.
\begin{equation} 
{R_{m_{i}n_{i}}}[g^{(i)}] = \lambda^{i} g^{(i)}_{m_{i}n_{i}},
\end{equation}
$i = 1, \ldots ,n $; $n \geq 2$.
The energy-momentum tensor is adopted in the following form
\begin{eqnarray} 
&& T^{M}_{N} = \sum_{\alpha = 0}^{m} T^{M (\alpha)}_{N}, \\
&&(T^{M (\alpha)}_{N})= diag(-{\rho^{(\alpha)}}(t),
 {p_{1}^{(\alpha)}}(t) \delta^{m_{1}}_{k_{1}},
\ldots , {p^{(\alpha)}_{n}}(t) \delta^{m_{n}}_{k_{n}}).
\end{eqnarray}
$\alpha = 0, \ldots ,m$, with the conservation law constraints imposed:
\begin{equation} 
\bigtriangledown_{M} T^{M (\alpha)}_{N}=0, \end{equation}
$\alpha= 0, \ldots ,m-1$.
The Einstein equations
\begin{equation}  
R^{M}_{N}-\frac{1}{2}\delta^{M}_{N}R=\kappa^{2}T^{M}_{N}
\end{equation}
($\kappa^{2}$ is gravitational constant) imply
$\bigtriangledown_{M} T^{M}_{N}=0$ and consequently
$\bigtriangledown_{M} T^{M (m)}_{N}=0$.

We suppose that for any $\alpha$-th component of matter the
pressures in all spaces are proportional to the density
\begin{equation} 
{p_{i}^{(\alpha)}}(t) = (1- \frac{u_{i}^{(\alpha)}}{N_{i}})
{\rho^{(\alpha)}}(t),
\end{equation}
where $u_{i}^{(\alpha)} = const$,
$i =1, \ldots ,n$;  $\alpha= 0, \ldots , m$.

Non-zero components of the Ricci-tensor for the metric
(2.1) are the following
\begin{equation} 
R_{00}=- \sum_{i=1}^{n} N_{i}[ \ddot{x}^{i} - \dot{\gamma} \dot{x}^{i}
+ (\dot{x}^{i})^{2}], \end{equation}
\begin{equation} 
R_{m_{i}n_{i}}=g_{m_{i}n_{i}}^{(i)} [\lambda^{i} +\exp(2x^{i}-2\gamma)
(\ddot{x}^{i}+\dot{x}^{i}(\sum_{i=1}^{n}N_{i}\dot{x}^{i}-\dot{\gamma}))],
\end{equation}
$i=1,\ldots,n$.

The conservation law constraint (2.6) for $\alpha \in \{0,...,m\}$ reads
\begin{equation} 
\dot{\rho}^{(\alpha)}+
\sum_{i= 0}^{n}N_{i}\dot{x}^{i}(\rho^{(\alpha)} + p_{i}^{(\alpha)})=0.
\end{equation}
 From eqs. (2.8), (2.11) we get
\begin{equation} 
{\rho^{(\alpha)}}(t)=A^{(\alpha)}
\exp[-2N_{i}{x^{i}}(t) + u_i^{(\alpha)}{x^i}(t)],
\end{equation}
where $A^{(\alpha)}=const$. Here and below the summation over
repeated indices is understood.

We define
\begin{equation} 
\gamma_{0} \equiv \sum_{i=1}^{n} N_{i}x^{i} \end{equation}
in (2.1).

Using relations (2.8), (2.9), (2.10), (2.12) it is not difficult to
verify that the Einstein equations (2.7) for the metric (2.1)
and the energy-momentum tensor from (2.4), (2.5) are
equivalent to the Lagrange equations for the Lagrangian
\begin{equation} 
L = \frac{1}{2} \exp(- \gamma + {\gamma_{0}}(x)) G_{ij}
\dot{x}^{i}\dot{x}^{j}- \exp( \gamma - {\gamma_{0}}(x)) {V}(x).
\end{equation}
Here
\begin{equation} 
G_{ij}=N_{i}\delta_{ij}- N_{i}N_{j} \end{equation}
are the components of the minisuperspace metric,
\begin{equation} 
V = {V}(x) = -\frac{1}{2}\sum_{i=1}^{n} \lambda^{i} N_{i}
\exp(-2x^{i}+2 {\gamma_{0}}(x)) +
\sum_{\alpha=0}^{m} \kappa^{2} A^{(\alpha)} \exp(u_i^{(\alpha)}x^i).
\end{equation}
is the potential. This relation  may be also presented in the form
\begin{equation} 
V = \sum_{\alpha=0}^{\bar{m}}  A_{\alpha} \exp(u_i^{(\alpha)}x^i),
\end{equation}
where  $\bar{m} = m + n$; $A_{\alpha} = \kappa^{2} A^{(\alpha)}$,
$\alpha= 0, \ldots , m$; $A_{m+i} =- \frac{1}{2} \lambda^{i} N_{i}$
and
\begin{equation} 
u_j^{(m+i)}= 2(- \delta^i_j + N_j),
\end{equation}
$i,j=1,\ldots,n$. We also put  $A_0 = \Lambda$ and
\begin{equation} 
u_j^{(0)}= 2 N_j,
\end{equation}
$j= 1,\ldots,n$. Thus the zero component of the matter
describe a cosmological constant term ($\Lambda$-term).

{\bf Diagonalization.} We remind [14,15] that the minisuperspace metric
\begin{equation} 
G = G_{ij}dx^{i} \otimes dx^{i}  \end{equation}
has a pseudo-Euclidean signature $(-,+, \ldots ,+)$, i.e. there exist
a linear transformation
\begin{equation} 
z^{a}=e^{a}_{i}x^{i},
\end{equation}
diagonalizing the minisuperspace metric (2.20)
\begin{equation} 
G= \eta_{ab}dz^{a} \otimes dz^{b}=
 -dz^{0} \otimes dz^{0} + \sum_{i=1}^{n-1}dz^{i}\otimes dz^{i},
\end{equation}
where
\begin{equation} 
(\eta_{ab})=(\eta^{ab}) \equiv diag(-1,+1, \ldots ,+1),
\end{equation}
$a,b = 0, \ldots ,n-1$. The matrix of the linear transformation
$(e^{a}_{i})$  satisfies the relation
\begin{equation} 
\eta_{ab}e^{a}_{i}e^{b}_{j} = G_{ij} \end{equation}
or equivalently
\begin{equation} 
\eta^{ab} = e^{a}_{i}G^{ij} e^{b}_{j} =  <e^{a},e^{b}>.
\end{equation}
Here
\begin{equation}  
G^{ij} = \frac{\delta^{ij}}{N_{i}}+ \frac{1}{2-D}
\end{equation}
are components of the matrix inverse to the matrix (2.15) [15],
$D = 1 + \sum_{i=1}^{n} N_i$ is the dimension of the manifold
$M$ (2.2) and
\begin{equation}  
<u,v> \equiv G^{ij}u_{i}v_{j}
\end{equation}
defines a bilinear form on $R^n$ ($u = (u_{i})$ , $v = (v_{i})$).
Inverting the map (2.21) we get
\begin{equation} 
x^{i} = e_{a}^{i} z^{a},
\end{equation}
where for the components of the inverse matrix
$(e_{a}^{i}) = (e^{a}_{i})^{-1}$ we obtain from (2.25)
\begin{equation} 
e_{a}^{i}    = G^{ij} e^{b}_{j} \eta_{ba}.
\end{equation}

Like in [15,21 ] we put
\begin{equation} 
z^0 = e^{0}_{i} x^i = q^{-1} N_i x^i, \qquad q = [(D-1)/(D-2)]^{1/2}.
\end{equation}
In this case the $00$-component of eq. (2.25) is satisfied and the set
$(e^a, a = 1, \ldots ,n-1)$ is defined up to
$O(n-1)$-transformation. A special example of the diagonalization with the
relations (2.30) and
\be  
z^a = e^{a}_{i} x^i = [N_a/ (\sum_{j=a}^{n} N_j)(\sum_{j=a+1}^{n} N_j)]^{1/2}
\sum_{j=a+1}^{n} N_j (x^j - x^i),
\ee
$a = 1, \ldots ,n-1$, was
considered in [14,15].

In $z$-coordinates (2.21) with $z^0$ from (2.30) the Lagrangian (2.14)
reads
\begin{equation} 
L = {L}(z^{a}, \dot{z}^{a}, {\cal N}) = \frac{1}{2} {\cal N}^{-1}
\eta_{ab} \dot{z}^{a} \dot{z}^{b} -  {\cal N} {V}(z),
\end{equation}
where
\begin{equation} 
{\cal N} = \exp( \gamma - {\gamma_{0}}(x)) > 0
\ee
is the Lagrange multiplier (modified lapse function) and
\begin{equation} 
{V}(z) =
\sum_{\alpha=0}^{\bar{m}} A_{\alpha} \exp(u^{\alpha}_a z^a)
\end{equation}
is the potential.
Here we denote
\begin{equation} 
u^{\alpha}_a  = e_{a}^{i} u^{(\alpha)}_i =
<u^{(\alpha)}, e^{b}> \eta_{ba},
\end{equation}
$a = 0, \ldots , n-1$, (see (2.27) and (2.29)). From (2.35) we get
(see (2.26), (2.27) and (2.30))
\begin{equation} 
u^{\alpha}_0  = - <u^{(\alpha)}, e^{0}> =
(\sum_{i=1}^{n} u^{(\alpha)}_i ) / q(D-2).
\end{equation}
For $\Lambda$-term and curvature components (see (2.19) and (2.18)) we
have
\begin{equation} 
u^0_0 = 2 q > 0 , \qquad u^{m + j}_0 = 2/q > 0,
\end{equation}
$j= 1,\ldots,n$. The calculation of
\begin{equation} 
(u^{\alpha})^2 = \eta^{ab} u^{\alpha}_a u^{\alpha}_b =
<u^{(\alpha)}, u^{(\alpha)}> = (u^{(\alpha)})^2 ,
\end{equation}
for these components gives
\begin{equation} 
(u^{0})^2 = 4(D-1)/(2-D) < 0, \qquad  (u^{m + j})^2 =
4(\frac{1}{N_j} - 1) < 0,
\end{equation}
for $N_j > 1$, $j= 1,\ldots,n$. For $N_j = 1$ we have $\lambda^{j}
= A_{m+j} = 0$.


\section{Billiard  representation}
\setcounter{equation}{0}

Here we consider the behavior of the dynamical system, described
by the Lagrangian (2.32) for $n \geq 3$ in the limit
\begin{equation} 
z^0  \rightarrow  -\infty, \qquad
z =(z^0, \vec{z}) \in {\cal V}_{-}, \end{equation}
where  ${\cal V}_{-}
\equiv \{(z^0, \vec{z}) \in R^n | z^0 < - |\vec{z}| \}$ is the lower light
cone. For the volume scale factor
\begin{equation} 
v = \exp(\sum_{i=1}^{n} N_{i}x^{i}) = \exp(qz^0)
\end{equation}
(see (2.30)) we have in this limit $v \rightarrow 0$.
Under certain additional assumptions the limit (3.1) describes
the approaching to the singularity.
We impose the following restrictions on the parameters $u^{\alpha}$
in the potential (2.34) for components with $A_{\alpha} \neq 0$:
\begin{eqnarray} 
&& 1) A_{\alpha} > 0  \  if \ (u^{\alpha})^2 = -(u^{\alpha}_0)^2 +
(\vec{u}^{\alpha})^2 > 0;         \\
&& 2) u^{\alpha}_0 > 0  \ for \ all \  \alpha.
\ea
We note that due to (2.37) the second condition is always
satisfied for $\Lambda$-term and curvature components (i.e. for
$\alpha = 0, m +1, \ldots, m+n = \bar{m}$).

We restrict the Lagrange system (2.32) on ${\cal V}_{-}$, i.e.
we consider the Lagrangian
\begin{equation} 
L_{-} \equiv L|_{TM_{-}} , \qquad M_{-} = {\cal V}_{-} \times R_{+},
\end{equation}
where $ TM_{-}$ is tangent vector bundle over $M_{-}$  and
$R_{+} \equiv \{ {\cal N} > 0 \}$. (Here $F|_{A}$ means the restriction
of function $F$ on $A$.)
Introducing  an analogue of the Misner-Chitre coordinates in
$\cal{V}_{-}$ [36,37]
\begin{eqnarray} 
&&z^0 = - \exp(-y^0) \frac{1 + \vec{y}^2}{1 - \vec{y}^2}, \\
&&\vec{z} = - 2 \exp(-y^0) \frac{ \vec{y}}{1 - \vec{y}^2},
\end{eqnarray}
$|\vec{y}| < 1$, we get for the Lagrangian (2.32)
\begin{equation} 
L_{-} = \frac{1}{2} {\cal N}^{-1} e^{- 2 y^0}
[- (\dot{y}^{0})^2 + {h_{ij}}(\vec{y}) \dot{y}^{i} \dot{y}^{j}]
-  {\cal N} V.
\end{equation}
Here
\be 
{h_{ij}}(\vec{y}) = 4 \delta_{ij} (1 - \vec{y}^2)^{-2},
\ee
$i,j =1, \ldots , n-1$, and
\be 
V = {V}(y) =
\sum_{\alpha=0}^{\bar{m}} A_{\alpha} \exp {\bar{\Phi}}(y,u^{\alpha}),
\end{equation}
where
\be 
{\bar{\Phi}}(y,u)  \equiv - e^{-y^0}(1 - \vec{y}^2)^{-1}
[u_0 (1 + \vec{y}^2) + 2 \vec{u}\vec{y}],
\ee

We note that the $(n-1)$-dimensional open unit disk (ball)
\be 
D^{n-1} \equiv \{ \vec{y}= (y^1, \ldots, y^n)| |\vec{y}| < 1 \}
\subset R^{n-1} \ee
with the metric $h = {h_{ij}}(\vec{y}) dy^i \otimes dy^j $ is one
of the realization of the $(n-1)$-dimensional Lobachevsky space
$H^{n-1}$.

We fix the gauge
\be 
{\cal N} =   \exp(- 2y^0) = - z^2.
\ee
Then, it is not difficult to verify that the
Lagrange equations for the Lagrangian (3.8) with the gauge fixing
(3.13)  are  equivalent  to  the Lagrange equations for the
Lagrangian
\be 
L_{*} = - \frac{1}{2}  (\dot{y}^{0})^2 +  \frac{1}{2}
{h_{ij}}(\vec{y}) \dot{y}^{i} \dot{y}^{j} -  V_{*}
\ee
with the energy constraint imposed
\be 
E_{*}
= - \frac{1}{2}  (\dot{y}^{0})^2 +  \frac{1}{2}
{h_{ij}}(\vec{y}) \dot{y}^{i} \dot{y}^{j} +  V_{*} = 0.
\ee
Here
\be 
V_{*} =  e^{-2y^0} V =
\sum_{\alpha=0}^{\bar{m}} A_{\alpha} \exp({\Phi}(y,u^{\alpha})), \ee
where
\be 
{\Phi}(y,u)  = - 2y^0 + {\bar{\Phi}}(y,u).
\ee

Now we are interested in the behavior of the dynamical system
in the limit  $y^0 \rightarrow - \infty$ (or, equivalently, in
the limit $z^2  = -(z^0)^2 + (\vec{z})^2
\rightarrow - \infty$, $z^0 < 0$) implying (3.1).
Using the relations  ($u_0 \neq 0$ )
\begin{eqnarray} 
&&{\Phi}(y,u) = - u_0 \exp(-y^0)
\frac{{A}(\vec{y}, -\vec{u}/u_0)}{1 - \vec{y}^2} - 2y^0, \\
&& {A}(\vec{y}, \vec{v}) \equiv
(\vec{y} - \vec{v})^2 -\vec{v}^2 + 1,
\end{eqnarray}
we get
\be 
\lim_{y^0 \rightarrow - \infty} \exp {\Phi}(y,u)  = 0
\ee
for $u^2 = - u_0^2 + (\vec{u})^2 \leq 0$, $u_0 > 0$ and
\be 
\lim_{y^0 \rightarrow - \infty} \exp {\Phi}(y,u)  =
{\theta_{\infty}}(-{A}(\vec{y}, - \vec{u}/u_0))
\ee
for $u^2 > 0$, $u_0 > 0$. In (3.21) we denote
\ba 
{\theta_{\infty}}(x) \equiv + &\infty, &x \geq 0,  \nonumber \\
                              & 0    , &x < 0.
\ea
Using restrictions (3.3), (3.4) and relations (3.16), (3.20),
(3.21) we obtain
\be 
{V_{\infty}}(\vec{y}) \equiv \lim_{y^0
\rightarrow - \infty} {V_{*}}(y^0, \vec{y}) = \sum_{\alpha \in
\Delta_{+}} {\theta_{\infty}}(-{A}(\vec{y}, -
\vec{u^{\alpha}}/u_0^{\alpha})).  \ee
Here we denote
\be 
\Delta_{+}  \equiv \{ \alpha | (u^{\alpha})^2 > 0 \}.  \ee
We note that due to (2.39) $\Lambda$-term and curvature components
do not contribute to $V_{\infty}$ (i.e. they  may be neglected
in the vicinity of the singularity).

The potential $V_{\infty}$
may be also written as following
\ba 
{V_{\infty}}(\vec{y}) =
{V}(\vec{y},B) \equiv &0, &\vec{y} \in B,
\nonumber \\
&+ \infty, &\vec{y}
\in D^{n-1} \setminus B,
\ea where
\be 
B = \bigcap_{\alpha \in
\Delta_{+}} {B}(u^{\alpha})  \subset D^{n-1}, \ee
\be 
{B}(u^{\alpha})  = \{ \vec{y} \in D^{n-1} |
|\vec{y} + \frac{\vec{u}^{\alpha}}{u_{0}^{\alpha}}| >
\sqrt{(\frac{\vec{u}^{\alpha}}{u_0^{\alpha}})^2 - 1} \},
\ee
$\alpha \in \Delta_{+}$. $B$ is an open domain.
Its boundary $\partial B = \bar{B} \setminus B$ is formed by
certain parts of $m_{+} = |\Delta_{+}|$
($m_{+}$ is the number of elements in $\Delta_{+}$) of $(n-2)$-dimensional
spheres with the centers in the points
\be 
\vec{v}^{\alpha} = - \vec{u}^{\alpha}/u^{\alpha}_{0}, \qquad
\alpha \in \Delta_{+},
\ee
($|\vec{v^{\alpha}}| > 1$) and radii
\be 
r_{\alpha} = \sqrt{(\vec{v}^{\alpha})^2 - 1}
\ee
respectively (for $n =3$, $m_{+} = 1$, see Fig. 1).

                                        Fig. 1

So, in the limit $y^{0} \rightarrow - \infty$ we are led to the
dynamical system
\ba
&L_{\infty} = - \frac{1}{2} (\dot{y}^{0})^2 +  \frac{1}{2}
{h_{ij}}(\vec{y}) \dot{y}^{i} \dot{y}^{j} -  {V_{\infty}}(\vec{y}), \\
&E_{\infty} = - \frac{1}{2} (\dot{y}^{0})^2 +  \frac{1}{2}
{h_{ij}}(\vec{y}) \dot{y}^{i} \dot{y}^{j} +  {V_{\infty}}(\vec{y}) = 0,
\ea
which after the separating of $y^0$ variable
\be
y^0 = \omega (t - t_0),
\ee
($\omega \neq 0$ , $t_0$  are constants) is reduced to the Lagrange
system with the Lagrangian
\be
L_{B} =  \frac{1}{2} {h_{ij}}(\vec{y})
\dot{y}^{i} \dot{y}^{j} -  {V}(\vec{y},B). \ee
Due to (3.32)
\be 
E_{B} =  \frac{1}{2}
{h_{ij}}(\vec{y}) \dot{y}^{i} \dot{y}^{j} +  {V}(\vec{y},B) =
\frac{\omega^2}{2}.
\ee
We put $\omega > 0$, then the limit $t \rightarrow - \infty$ describes
the approach to the singularity.
When the set (3.24) is empty ($\Delta_{+} = \emptyset$)  we have $B = D^{n-1}$
and the Lagrangian (3.33)  describes  the  geodesic  flow  on  the
Lobachevsky space $H^{n-1} = (D^{n-1}, h_{ij} dy^i \otimes dy^j)$. In
this case there are two families of non-trivial geodesic solutions
(i.e. ${y}(t) \neq const$):
\ba
1. &{\vec{y}}(t) = \vec{n}_1 [\sqrt{v^2 -1} \cos {\varphi}(\bar{t}) - v]
                 +  \vec{n}_2 \sqrt{v^2 -1} \sin {\varphi}(\bar{t}) ,  \\
   &{\varphi}(\bar{t})  = 2 \arctan
   [(v - \sqrt{v^2 -1}) \tanh (\omega \bar{t})], \\
2. &{\vec{y}}(t) = \vec{n} \tanh (\omega \bar{t}).
\ea
Here $\vec{n}^2 = \vec{n}_1^2 = \vec{n}_2^2 =1$, $\vec{n}_1 \vec{n}_2
= 0$, $v > 1$, $\omega > 0$, $\bar{t} = t - t_0$, $t_0 = const$.

Graphically the first solution corresponds to the arc of the circle with
the center at point ($-v \vec{n}_1$) and the radius $\sqrt{v^2 -1}$.
This circle belongs to the plane spanned by vectors
$\vec{n}_1$ and  $\vec{n}_2$
(the centers of the circle and the ball $D^{n-1}$ also belong
to this plane). We note, that the solution (3.35)-(3.36) in the limit
$v \rightarrow \infty$ coincides with the solution (3.37).

We note, that the boundary of the billiard $\partial B$ is formed
by geodesics. For some billiards  this fact  may be used
for  "gluing" certain parts of boundaries.

When  $\Delta_{+} \neq \emptyset$ the Lagrangian
(3.33) describes the motion of the particle  of  unit  mass,  moving  in
the ($n-1$)-dimensional billiard $B \subset D^{n-1}$  (see (3.26)).
The geodesic motion in
$B$ (3.35)-(3.37) corresponds to a "Kasner epoch" and the reflection from
the boundary corresponds to the change of Kasner epochs.  For $n = 3$
some examples  of (2-dimensional) billiards are depicted in Figs. 2-4.

\begin{center}   Figs. 2-4             \end{center}

The billiard $B$ in Fig. 2. has an infinite volume: $vol B = +\infty$.
In this case there are three open zones  at  the  infinite  circle
$|\vec{y}| =1$. After a finite number of reflections from the boundary
the  particle  moves  toward  one  of  these  open  zones.  For
corresponding cosmological model we get the "Kasner-like"
behavior in the limit $t \rightarrow - \infty$ [18].

For billiards depicted in Figs. 3 and 4 we have $vol B < + \infty$.
In the first case (Fig. 3) the
closure of the billiard  $\bar{B}$                                is
compact (in the topology of $D^{n-1}$) and in the
second case (Fig. 4) $\bar{B}$ is non-compact. In these two cases the
motion of the particle is stochastic.

Analogous arguments may be applied to the to the case $n > 3$.
So, we are interested  in  the configurations with finite volume of $B$.
We propose a simple geometric criterion  for the
finiteness of the volume of $B$ and compactness of $\bar{B}$
in terms of the positions of the
points (3.28)  with respect to the ($n-2$)-dimensional unit sphere $S^{n-2}$
($n \geq 3$).  We say that the point $\vec{y} \in S^{n-2}$ is (geometrically)
illuminated by the point-like source located at the point  $\vec{v}$,
$|\vec{v}| > 1$,  if and only if $|\vec{y} - \vec{v}| \leq
\sqrt{|\vec{v}|^2 -1}$. In Fig. 1 the source $P$ illuminates the closed
arc $[P_1,P_2]$.  We also say that the point $\vec{y} \in S^{n-2}$ is
strongly illuminated by the point-like source located at the point
$\vec{v}$, $|\vec{v}| > 1$, if and only if $|\vec{y} - \vec{v}| <
\sqrt{|\vec{v}|^2 -1}$. In Fig. 1 the source $P$ strongly illuminates the
open arc $(P_1,P_2)$. The subset $N \subset S^{n-2}$  is called
(strongly) illuminated by point-like sources at $\{ \vec{v}^{\alpha}, \alpha
\in \Delta_{+} \}$  if and only if any point from $N$ is (strongly)
illuminated by some source at $\vec{v}^{\alpha}$ ($\alpha \in \Delta_{+}$).

Proposition.
The billiard $B$ (3.26) has a finite volume if and only if the point-like
sources of light located at the points $\vec{v}^{\alpha}$ (3.28)
illuminate the unit sphere $S^{n-2}$.
The closure of the billiard  $\bar{B}$ is compact
(in the topology of $D^{n-1} \simeq H^{n-1}$) if and only if
the sources at points (3.28) strongly illuminate $S^{n-2}$.

Proof. We consider the set $\partial^c B  \equiv B^c \setminus \bar{B}$,
where $B^{c}$ is the completion of $B$ (or, equivalently, the
closure of $B$ in the topology
of $R^{n-1}$). We remind that $\bar{B}$ is the closure of $B$ in the
topology of $D^{n-1}$. Clearly, that $\partial^{c} B$ is a closed subset
of $S^{n-2}$, consisting of all those points that are not strongly
illuminated by sources (3.28). There are three possibilities:  i) $\partial^c
B$ is empty; ii) $\partial^{c} B$ contains some interior point (i.e. the
point belonging to $\partial^{c} B$ with some open neighborhood); iii)
$\partial^{c} B$ is non-empty finite set, i.e. $\partial^{c} B  = \{
\vec{y}_1, \ldots \vec{y}_l \}$. The first case i) takes place if and only
if $\bar{B}$ is  compact in the topology of $D^{n-1}$. Only in this case
the sphere $S^{n-2}$ is strongly illuminated by the sources (3.28). Thus
the second part of proposition is proved. In the case i) $vol B$ is
finite. For the volume we have
\be 
vol B = \int_{B} d^{n-1} \vec{y}
\sqrt{h} = \int_{0}^1 dr (1- r^2)^{1-n} S_{r}.  \ee
The "area" $S_{r}
\rightarrow C > 0$ as $r \rightarrow 1$ in the case ii) and, hence, the
integral (3.38) is divergent.  In the case iii)
\be 
S_{r} \sim C_1
(1 - r)^{2(n-2)}  \ as  \   r \rightarrow 1
\ee
($C_1 >0$) and, so, the
integral (3.38) is  convergent.  Indeed, in the case iii), when  $r
\rightarrow 1$,  the "area" $S_{r}$ is the sum of $l$ terms. Each of these
terms is the $(n-2)$-dimensional "area" of a transverse side of a deformed
pyramid with a top at some point $\vec{y}_k$, $k =1, \ldots , l$. This
multidimensional pyramid is formed by certain parts of spheres orthogonal
to $S^{n-2}$ in the point of their intersection $\vec{y}_k$. Hence, all
lengths of the transverse section  $r = const$ of the "pyramid" behaves
like $(1 - r)^{2}$, when  $r \rightarrow 1$,  that justifies (3.39).
But the unit sphere $S^{n-2}$ is illuminated
by the sources (3.28) only in the cases i) and iii). This completes the
proof.

The problem of illumination of convex body in
multidimensional vector space by point-like sources for the first time was
considered in [45,46]. For the case of $S^{n-2}$ this problem is equivalent
to the problem of covering the spheres with spheres [47,48]. There exist a
topological bound on the number of point-like sources $m_{+}$ illuminating
the sphere $S^{n-2}$ [46]:
\be 
m_{+} \geq n.
\ee

Remark 1. Let the points (3.28) form an open convex polyhedron
$P \subset R^{n-1}$. Then the sources at (3.28)
illuminate $S^{n-2}$, if $D^{n-1} \subset P$, and strongly
illuminate $S^{n-2}$, if $\overline{D^{n-1}} \subset P$.

{\bf Scalar field generalization}. Let us assume that an
additional $(m+1)$-th component with the equation of state
$p_{i}^{(m+1)} = \rho^{(m+1)}$ is considered, $i =1, \ldots ,n$.
This component  describes Zeldovich matter [49] in all spaces and  is
equivalent to homogeneous massless free minimally coupled scalar field [50].
In this case $u_{i}^{(m+1)} = 0$, $i =1, \ldots ,n$ and the potential
(2.17) is modified by the addition of constant $A_{m+1} > 0$. Then
the potential $V_{*}$ (3.16) is modified by the addition of
the following term
\be
\Delta V = A_{m+1} \exp(-2 y^0).
\ee
This do not prevent from the formation of the billiard walls but change the
 time dependence of $y^0$-variable:
\be
\exp(2 y^0) = 2 A_{m+1} \sinh^2[\omega (t - t_0)] / \omega^2,
\ee
($\omega > 0$) instead of (3.32). In the limit $t \rightarrow t_0 + 0$
we have $y^0 \rightarrow - \infty$ and  ${\vec{y}}(t)
\rightarrow \vec{y}_0 \in B$. So, the stochastic behavior near the
singularity is absent in this case.

\section{Bianchi-IX cosmology}
\setcounter{equation}{0}

Here we  consider the well-known mixmaster
model [34,35] with the metric
\begin{equation}  
g = - \exp[2{\gamma}(t)] dt \otimes dt +
\sum_{i=1}^{3} \exp[2{x^{i}}(t)] e^i \otimes e^i,
\end{equation}
where 1-forms  $e^i = {e^i_{\nu}}(\zeta) d \zeta^{\nu}$ satisfy the
relations
\be 
de^i = \frac{1}{2} \varepsilon_{ijk} e^j \wedge e^k,
\ee
$i,j,k = 1,2,3$. The Einstein equations for the metric (4.1)
lead to the  Lagrange system
(2.14)-(2.17) with (see, for example, [35])
$n =3$, $N_1 = N_2 = N_3 = 1$, $m=6$,
$A_{1}=A_{2}= A_{3}= 1/4$, $A_{4}=A_{5}= A_{6}= - 1/2$,
$A_{0}=A_{7}= A_{8}= A_{9} = 0$, and
\be 
u^{(\alpha)}_i = 4 \delta^{\alpha}_i, \qquad
u^{(3 + \alpha)}_i = 2 (1 - \delta^{\alpha}_i),
\ee
$\alpha = 1,2,3$.
In the $z$-coordinates (2.30), (2.31) we have for 3-vectors (2.35)
\be 
u^{1} = \frac{4}{\sqrt{6}} (1,1, - \sqrt{3}), \
u^{2} = \frac{4}{\sqrt{6}} (1,1, + \sqrt{3}), \
u^{3} = \frac{4}{\sqrt{6}} (1,-2,0),
\ee
\be 
u^{4} = \frac{1}{2}(u^{1} + u^{2}), \ u^{5} = \frac{1}{2}(u^{1} + u^{3}), \
u^{6} = \frac{1}{2}(u^{2} + u^{3}),
\ee
and, consequently,
\be 
(u^{\alpha})^2  = 8, \qquad
(u^{3 + \alpha})^2 = 0,
\ee
$\alpha = 1,2,3$.
Thus the conditions (3.3), (3.4) are satisfied. The components
with $\alpha = 4,5,6$ do not survive in the approaching to the singularity .
For the vectors (3.28) we have
\be 
\vec{v}^1 = (1, - \sqrt{3}), \ \vec{v}^2 = (1, + \sqrt{3}),  \
\vec{v}^3 = (-2, 0),
\ee
i.e. a triangle from Fig. 4 (see also [38]). In this case the circle
$S^1$ is illuminated by sources at points $\vec{v}^i$,
$i= 1,2,3$, but not strongly illuminated. In agreement with
Proposition the billiard $B$ has finite volume, but $\bar{B}$
is not compact.

\section{Quantum case}
\setcounter{equation}{0}

The quantization of zero-energy constraint (3.15) leads to the
Wheeler-DeWitt (WDW) equation in the gauge (3.13) [15,52]
\begin{equation} 
(- \frac{1}{2} {\Delta}[\bar{G}]  + a_{n}
{R}[\bar{G}]  + V_{*}) \Psi =0.
\end{equation}
Here
$\Psi={\Psi}(y)$ is "the wave function of the Universe",
$V_{*} = {V_{*}}(y)$ is the
potential (3.16), $a_{n} = (n-2)/8(n-1)$,
${\Delta}[\bar{G}]$ and ${R}[\bar{G}]$ are the Laplace-Beltrami
operator and the scalar curvature of the minisuperspace metric
\begin{equation} 
\bar{G} = - dy^0 \otimes dy^0 + h, \qquad
 h = {h_{ij}}(\vec{y}) dy^i \otimes dy^j.
\end{equation}
(We remind that, $h$ is the metric on Lobachevsky space $D^{n-1}$.)
The form of WDW eq. (5.1) follows from the demands of minisuperspace
invariance and conformal covariance [51,52,15]. Using relations
\begin{equation} 
{\Delta}[\bar{G}]  = - (\frac{\partial}{\partial y^0})^2 + {\Delta}[h],
\qquad {R}[\bar{G}] =   {R}[h] = - (n-1)(n-2),
\ee
we rewrite (5.1) in the form
\begin{equation} 
(\frac{1}{2} (\frac{\partial}{\partial y^0})^2 - \frac{1}{2} {\Delta}[h] -
\frac{(n-2)^2}{8}  + V_{*}) \Psi =0.
\end{equation} In the limit $y^0
\rightarrow - \infty$ the WDW eq. reduces to the relations
\begin{equation} 
((\frac{\partial}{\partial y^0})^2 - {\Delta_{*}}[h]) \Psi_{\infty} = 0,
\qquad \Psi_{\infty}|_{\partial B} = 0,
\ee where $\partial B = \bar B
\setminus B$ is the boundary of the billiard $B$ (3.26) (in $D^{n-1}$) and
\begin{equation} 
{\Delta_{*}}[h] =  {\Delta}[h] + \frac{(n-2)^2}{4}.
\ee
Now, we suppose that $\bar{B}$ is compact and the operator (5.6)
with the boundary condition (5.5) has a negative spectrum, i.e.
\begin{equation} 
{\Delta_{*}}[h] \Psi_{n} = -E_{n}^2  \Psi_{n},
\ee
$E_n > 0$, $n = 0,1, \ldots$ (this is valid at least for "small
enough" $B$). Using (5.7) we get the general solution
of the asymptotic WDW eq. (5.5)
\begin{equation} 
{\Psi_{\infty}}(y^0, \vec{y}) = \sum_{n=0}^{\infty}
[c_n \exp(-i E_n y^0) {\Psi_{n}}(\vec{y}) +
c_n^{*} \exp(i E_n y^0) {\Psi_{n}^{*}}(\vec{y})],
\ee
that may be considered as a starting point for the
construction of third quantized models in the vicinity
of the singularity.

\section{Discussions}
\setcounter{equation}{0}

Thus, we obtained the "billiard  representation"
for the cosmological model [21] and proved the  geometrical criterion
for the finiteness of the billiard volume and the compactness of the billiard
(Proposition, Sec. 3). This criterion may be used as a rather effective
(and universal) tool for the selection of the cosmological models with a
stochastic behavior near the singularity.

For an "isotropic" component: $p_{i}^{(\alpha)} = (1 - h) \rho^{(\alpha)}$,
$i = 1, \ldots , n$, with $h \neq 0$ we have $(u^{(\alpha)})^2 =
h^2 (D-1)/(2-D) < 0$ and, hence, this component may be neglected near
the singularity. Only "anisotropic" components with $(u^{(\alpha)})^2  > 0$
take part in the formation of billiard walls near the singularity.
According to the topological bound (3.40) [46] the stochastic behavior
near the singularity in the considered model may occur only if the
number of components with  $(u^{(\alpha)})^2  > 0$ is not less than the
minisuperspace dimension.

We also note that here, like  in the Bianchi-IX case [36,37], the considered
reduction scheme uses a special time gauge (or parametrization of time).
As it was pointed in [38] one should be careful in the interpretations of
the results of computer experiments for other choices of time.

{\bf Restrictions on parameters.} Here we discuss the physical sense of
the restrictions on parameters of the model (3.3) and (3.4). The
condition (3.3) means that the  densities of the
"anisotropic" components with  $(u^{(\alpha)})^2  > 0$
should be positive. Using
(2.8) and (2.36) we rewrite the  restriction (3.4) in the equivalent
form
\begin{equation} 
\sum_{i=1}^{n} N_{i} \frac{\rho^{(\alpha)} - p_{i}^{(\alpha)}}
{\rho^{(\alpha)}} > 0,
\end{equation}
($\rho^{(\alpha)} \neq 0$) $\alpha = 1,
\ldots , m$ (for curvature and $\Lambda$-terms (3.4) is satisfied). For
\begin{equation} 
\rho^{(\alpha)} > 0, \qquad p_{i}^{(\alpha)} < \rho^{(\alpha)},
\end{equation}
$\alpha = 1, \ldots , m$, $i = 1, \ldots , n$, (6.1) is satisfied
identically.

Remark 2. It may be shown [53] that the condition (3.4) may be weakened by
the following one
\be 
 u^{\alpha}_0 > 0, \  if  \ (u^{\alpha})^2 \leq 0.
\ee
In this case there exists a certain generalization of the set
${B}(u^{\alpha})$ from (3.27) for arbitrary $u^{\alpha}_0$  ($(u^{\alpha})^2
> 0$) [53].  The Proposition (Sec. 3) should be modified by including into
consideration the sources at infinity (for $u^{\alpha}_0 = 0$) and
"anti-sources" (for $u^{\alpha}_0 < 0$). For "anti-source" the shadowed
domain coincides with the illuminated domain for the usual source (with
$u^{\alpha}_0 > 0$).  In this case we deal with  the  kinematics of tachyons.
(We may also consider a covariant and slightly more general
condition instead of (6.3)
\be 
{\rm sign}u^{\alpha}_0 = \varepsilon,
\  for  \  all  \  (u^{\alpha})^2 \leq 0,  \  \varepsilon = \pm 1.)
\ee

\begin{center} {\bf Acknowledgments}   \end{center}

The authors would like to thank our colleagues K.A.Bronnikov,
A.A.Kirillov, M.Yu.Konstatinov and  A.G.Radynov for useful discussions.
One of us (V.D.I) is grateful to R.V.Galiullin for pointing out the
attention to ref. [46].

This work was supported in part by the Russian Ministry of Science.

\pagebreak

\pagebreak

\begin{center} {\bf List of captions for illustrations}
\end{center}

Fig. 1. An example of billiard for $n=3$, $m_{+} = 1$.

Figs. 2. Billiard with infinite volume for $n=3$, $m_{+} = 3$.

Figs. 3. Compact billiard for $n=3$, $m_{+} = 3$.

Fig. 4. Billiard corresponding to Bianchi-IX model
(non-compact with infinite volume).

\end{document}